\documentclass[
aps,
bm,
showpacs,
reprint]{revtex4-1}

\usepackage{amsmath,amssymb, amsfonts}
\usepackage[usenames,dvipsnames]{color}
\usepackage{dcolumn}

\usepackage{mathrsfs}  
\usepackage{graphicx}   
\usepackage{epstopdf}
\usepackage{color}

\def\bs{\begin{subequations}}
\def\es{\end{subequations}}
\def\aa{\begin{align}}
\def\ab{\end{align}}
\def\ba{\begin{eqnarray}}
\def\ea{\end{eqnarray}}
\def\be{\begin{equation}}
\def\ee{\end{equation}}

\begin{document}


\title{Elastic Platonic Shells}

\author{Ee Hou Yong$^1$, David R. Nelson$^{1, 2}$ and L. Mahadevan$^{1,2}$}
\affiliation{$^1$ Department of Physics, Harvard University, Cambridge, Massachusetts 02138\\
$^2$ School of Engineering and Applied Sciences, Harvard University, Cambridge, MA 02138}
\date{\today}

\begin{abstract}
On microscopic scales, the crystallinity of flexible tethered or cross linked membranes determines their mechanical response. We show that by controlling  the type, number and distribution of defects on a spherical elastic shell, it is possible to direct the morphology of these structures. Our numerical simulations show that by deflating a crystalline shell with defects, we can create elastic shell analogs of the classical Platonic solids. These morphologies arise via a sharp buckling transition from the sphere which is strongly hysteretic in loading-unloading. We construct a minimal Landau theory for the transition using quadratic and cubic invariants of the spherical harmonic modes. Our approach suggests methods to engineer shape into soft spherical shells using a frozen defect topology.  
\end{abstract}


\maketitle
The continuum theory of elastic shells is applicable to the study of the mechanical response of systems across a wide range of length scales, from viruses (25nm), vesicles (1$\mu$m) \cite{Taniguchi08, Pleiner10}, pollen grains (10-100$\mu$m) \cite{Eleni10}, armored bubbles (10-100$\mu$m)\cite{Stone07} to the behavior of macroscopic shells seen in aircraft fuselages (10m) and even megascale lithospheric dynamics \cite{Maha2010}. While the only geometric parameter in the study of these systems is their size to thickness ratio (which also translates into the only material parameter as well), on microscopic scales, the effects of crystallinity and defects may be important in determining the mechanical response of these shells. Indeed, isolated 5-fold disclinations in flexible membranes with internal crystalline order are responsible for the buckling of a flat membrane \cite{Seung88}, an effect that manifests itself in the distinctive icosahedral structure of virus capsids (100nm) \cite{Caspar62, Nelson03, Nelson07}, the shape of colloidal shells \cite{Datta2010} etc. To study these systems where crystallinity is potentially important, we need to account for the dynamics of the defects while simultaneously following the dynamics of the embedding shell, which might engender new defects. This is a difficult task in general, and so most approaches decouple these two different processes based on the rate at which they happen. The two extreme limits correspond to the cases when the topography is frozen, but the defects are mobile \cite{Bowick00, Bowick07, Luca07}, and the case when then the defects are frozen but the topography is mobile \cite {Nelson03, Nelson07}. We focus here on the latter case,  and show that by playing with the number, type and arrangement of defects on a soft spherical shell, and then deflating it, we can derive controllable morphologies that resemble the Platonic solids.  

\begin{figure}[htbp]
\centering
\includegraphics[width=8.6cm]{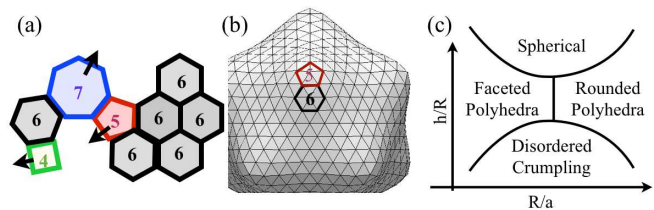}
\caption{\label{fig:intro} (Color online) Difference between frozen topography and frozen defects. (a) Frozen topography: There is a 4-fold disclination (green square), a 5-fold disclination (red pentagon) and a 7-fold disclination (blue heptagon) that are free to move in a ``sea" of 6-fold sites (grey hexagons), often at the expense of creating additional defects \cite{Bowick07}. (b) Frozen defects: The defects are frozen and the overall geometry (shape) of the surface can change. (c) Phase diagram of buckled membranes as function of $h/R$ and $R/a$. }
\end{figure}

For crystalline complete spherical shells, topological considerations pose constraints that dictate that the number and type of disclinations must satisfy the condition known as Euler's formula, $\sum_z (6 - z) N_z = \sum_z q_z N_z = 12$ \cite{supmat}, where $N_z$ is the number of vertices with $z$-coordination number and $q_z = 6 - z$ is the topological charge of a vertex, as shown in the examples in Fig.~\ref{fig:intro}a. We assume that the core energies of disclinations are large so that a crystalline shell prefers to have the minimum number of isolated disclinations that satisfy Euler's formula and denote the number of 3-, 4-, 5-fold disclinations by $\vec{n} = \{n_3, n_4, n_5\}$, e.g. $\{ 0, 6, 0\}$ refers to a shell with 6 4-fold disclinations; a regular octahedron is an especially simple example. There are a total of 19 different possibilities that satisfy Euler's formula and they fall into 3 distinct universality classes \cite{supmat}. We will further assume that the set of disclinations obeys some group symmetry $G$, i.e., the set of topological defects are invariant under the action of group $G$. Then, $\vec{n}$ and $G$ maps any spherical surface with defects onto a unique polyhedron (many-to-one map) and hencefore, we will identify each surface with the corresponding polyhedron. The shape of the deformed shell depends sensitively on the thickness of the shell $h$, the radius of the shell $R$ and the average vertex spacing $a$ from which we can form two dimensionless parameters $h/R$ (aspect ratio), which characterizes the slenderness of the shell and $R/a$ (lattice ratio) which characterizes the discreteness of the shell. 

We drive the formation of the polyhedral morphologies of these thin spherical shells by deflating them gradually \cite{Siber06}. This naturally leads to shapes that minimize the total energy of a thin two-dimensional shell that is the sum of the stretching energy and the bending energy, i.e., $U_T = U_S + U_B$ \cite{Landau86}. Due to the slenderness of the shell, it is energetically favorable to bend rather than to stretch, resulting in a highly faceted shape \cite{Witten07} whence the energy, which is initially smoothly distributed, becomes more and more non-uniform, with high energy concentrated in the bent regions (edges and vertices). Owing to the geometrical nonlinearity of the resulting energy densities, we used a numerical approach to determine the morphologies using Surface Evolver \cite{Brakke92}. We construct a spherical shell with crystalline order containing a certain set of fixed defects of different types that satisfy Euler's formula and then decrease the volume in small decrements and equilibrate the elastic energy in each step. We find that beyond a critical decrement in the volume, the shells buckles into different faceted shapes such as that shown in Fig.~\ref{fig:intro}b; the specific form is constrained by the number, type and orientation of the defects in the original spherical shell. As we vary the aspect ratio $h/R$, we find that thick spheres ($h/R \gtrsim 0.1$) deform isotropically and the shell is always smooth with no noticeable faceting; thin spherical shells tend to buckle into highly faceted structures. On the other hand, at low lattice ratio $R/a$, we generally get simpler buckled structures since there are fewer degrees of freedom; as we increase $R/a$, we get more complicated structures. At intermediate values of $h/R$ and $R/a$, we get structures that resemble regular polyhedra. In general, we expect to see various morphologies as a function of the two geometrical parameters corresponding to the aspect ratio and the lattice ratio as schematized in Fig.~\ref{fig:intro}c. 
\begin{figure}[htbp]
\centering
\includegraphics[width=8.0cm]{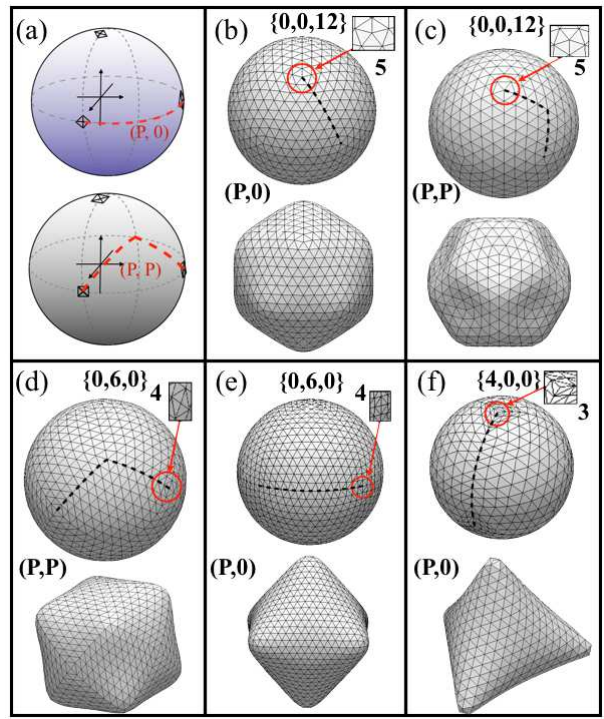}
\caption{(Color online) (a) Difference between $(P,0)$ and $(P,P)$ shells illustrating importance of the orientation of defects relative to the crystallographic axes. (b)--(d) Simulations of crystalline shells with different topological defects. Black dotted line shows path between two disclinations and number indicate coordination number. Top (bottom) panel of each box denotes initial (final) state. Final state is at volume fraction $\approx0.8$. (b) Icosahedron-shell. (c) Dodecahedron-shell. (d) Cube-shell. (e) Octahedron-shell. (f) Tetrahedron-shell.}
\label{fig:shape}
\end{figure} 

In Fig.~\ref{fig:shape}, we see that all shells corresponding to Platonic solids arise as a function of the nature, number and location of the defects on a sphere. Limiting ourselves to the case of a single class of defects, we use the classical Caspar-Klug notation $(P,Q)$ \cite{Caspar62}, where $P, Q \in \mathbb{N}$. Thus, for $\vec{n}=\{0,0,12\}$ and $(P,0)$, we get an icosahedron-shell; for $\vec{n}=\{0,0,12\}$ and $(P,P)$, we get its dual dodecahedron-shell; for $\vec{n}=\{0,6,0\}$ and $(P,0)$, we get an octahedron-shell; for $\vec{n}=\{0,6,0\}$ and $(P,P)$ we get its dual cube-shell and finally $\vec{n}=\{4,0,0\}$ and $(P,0)$, yields a tetrahedron-shell. We can analyze the shapes of the buckled shells quantitatively by looking at the spherical harmonic expansion of the shape as characterized by the position of the vertices \cite{Nelson81, Nelson83, Nelson03}
\ba
D(\theta, \phi) =&& \sum_{i=1}^N R_i \delta(\phi - \phi_i)\delta(\cos \theta - \cos \theta_i) \nonumber \\
\approx&& \sum_{\ell = 0}^{L} \sum_{m = -\ell}^{\ell} a_{\ell}^{m} Y_{\ell}^{m}(\theta, \phi),
\ea
where $(\theta_i, \phi_i, R_i)$ represents the polar coordinates of vertex $i$ ($i = 1, \dots N$). Since $a_{\ell}^{m}$'s are coordinate-dependent, we consider rotationally invariant quantities formed from $a_{\ell}^{m}$'s that measure the angular projection onto the different $\ell$'s \cite{Nelson81, Nelson83}. Two such quantities are
\be
Q_{\ell} = \frac{1}{a^0_0}\left( \frac{1}{2\ell +1} \sum_{m = -\ell}^{\ell} |a_{\ell}^m|^2\right)^{1/2},
\ee
and
\be 
W_{\ell}  = \frac{\sum_{\Omega} \left( 
\begin{array}{ccc}
\ell & \ell & \ell \\ 
m_1 & m_2 & m_3 
\end{array} \right) a_{\ell}^{m_1} a_{\ell}^{m_2} a_{\ell}^{m_3}}{\left(\sum_m |a_{\ell}^{m} |^2\right)^{3/2}},
\label{eq:wl}
\ee
where $\Omega$ denotes the set of $m$'s such that  $m_1+m_2+m_3 = 0$ and the parenthesis term in Eq.~(\ref{eq:wl}) is the Wigner $3j$-symbol. These parameters allow us to carry out a ``shape spectroscopy". For example, $Q_{\ell}$ with $\ell > 0$ measures the aspherity of the shell; $Q_{4}$ measures tetrahedral or cubic symmetry and $Q_{6}$ is an icosahedral order parameter \cite{Busse75, Sattinger78,Nelson83} while $W_{\ell}$ measures the orientational symmetry type of the buckled membrane; $W_4$ and $W_8$ vanish for icosahedra and dodecahedra, and have different relative weights for the other Platonic solids. They are normalized such that their magnitude is invariant to overall rescaling in $a_\ell^m$'s and $Q_0 = W_0 = 1$. In general, for a crystalline shell, we can evaluate its orientational symmetry by evaluating $Q_{\ell}$/$W_{\ell}$'s and then compare them with the $Q_{\ell}$/$W_{\ell}$'s of the corresponding Platonic solid, for which the spherical harmonic representation can be calculated using known algebraic formulae \cite{Onaka06}. The initial crystalline shell has nearly perfect spherical symmetry, i.e. $Q_\ell \approx \delta_{\ell,0}$. However, as the membrane buckles, the deformed shell starts to take on interesting shapes, with nonvanishing $Q_{\ell}$'s for $\ell > 0$. We calculate the $Q_{\ell}$'s for the various buckled crystalline shells in Fig~\ref{fig:shape}. The buckled icosahedron-shell and dodecahedron-shell have nonzero spherical harmonics only for $\ell = 0, 6, 10, 12, ...$; the cube-shell and octahedron-shell have nonvanishing spherical harmonics for $\ell = 0, 4, 6, 8, 10, ...$; the tetrahedron-shell has $\ell =  0, 3, 4, 6, 7, 8, 9, 10, ...$.  

\begin{table}[htbp]
\caption{\label{tab:WL} Normalized invariant $W_{\ell}$ for the Platonic solids.}
\begin{tabular}{@{}l|c|c|c|c}
\hline  
Type & $W_{4}$ & $W_{6}$ & $W_{8}$ & $W_{10}$\\
\hline  
Icosahedron & -- & $-$0.1697 & -- & +0.0940 \\
Dodecahedron & -- & $+0.1697$ & -- & $-0.0940$ \\
Cube & $-$0.1593 & +0.0132 & +0.0584 & $-$0.0901 \\
Octahedron & +0.1593 & $-$0.0132 & +0.0584 & $-$0.0901  \\
Tetrahedron & +0.1593 & +0.0132 & +0.0584 & $-$0.0901 \\
\hline  
\end{tabular}
\end{table}

Despite their rather different shapes, we find that the icosahedron-shell and dodecahedron-shell have identical magnitudes of $W_{\ell}$'s, as they belong to the same symmetry group $G$ and these parameters characterize the symmetry of the vertices. We also compare the set of $|W_{\ell}|$'s of the buckled shells with their values for the ideal Platonic solids as shown in Table~\ref{tab:WL}. In the case of the icosahedron-shell and dodecahedron-shell, we find that the first two non-zero $W_{\ell}$'s are $|W_{6}| = 0.1697$ and $|W_{10}| = 0.0940$; furthermore $(a_6^0)^2 = \frac{11}{7} |a_{6}^{\pm5}|^2$, with all the other $a_{6}^{m}$ vanishing, coefficients which maximize $|W_6|$ \cite{Busse75, Sattinger78, Nelson83}. Similarly, for the cube-shell and octahedron-shell, we find that $(a_4^0)^2 = \frac{14}{5} |a_{4}^{\pm4}|^2,$ with all other $a_4^m = 0$, coefficients which maximize $|W_4|$ \cite{Busse75, Sattinger78, Nelson83}. For the self-dual tetrahedron $W_3$ vanishes even though $Q_3 \neq 0$ \cite{foot1}.  Our analysis of the numerical simulations shows that the simplest shape parameters $Q_{\ell}$ and $W_{\ell}$ for the 5 Platonic-shells converge to that of the actual Platonic solids. Thus, as long as the crystalline shell has frozen defects, this symmetry leads to buckled shapes with the same symmetry.

\begin{figure}[htbp]
\begin{center}
\includegraphics[width=8.6cm]{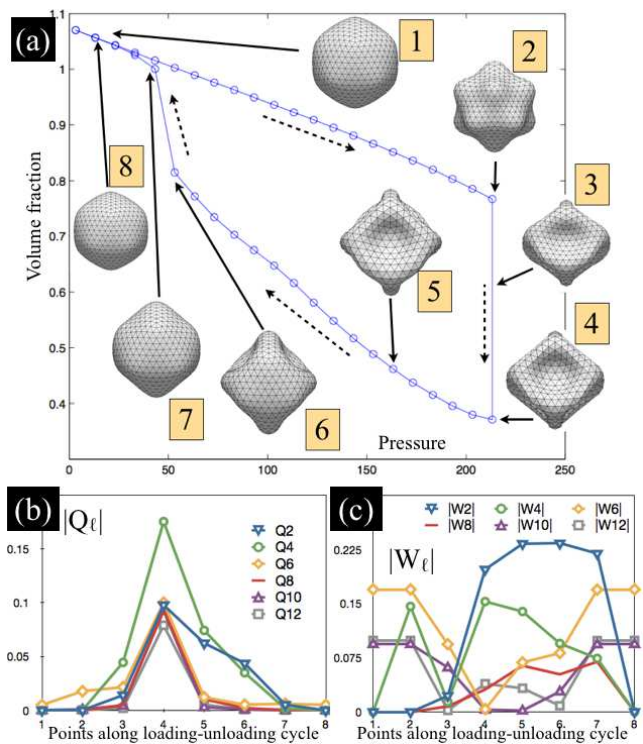}
\end{center}
\caption{(Color online) (a) Mechanical response of a crushed icosahedral-shell during loading-unloading (LU) cycle. The labels $(1)$ to $(8)$ show the shape of the shell at different points during LU cycle. As the pressure is gradually increased, the shell becomes more undulating as reflected in (2). At the upper critical buckling pressure $p_b^u \approx 210$, the icosahedral-like shell undergoes an abrupt collapse into a vastly different looking shell with an approximately cubic symmetry in (4). As the pressure is decreased, the shell does not return to the inflated condition by the same path at the pressure-volume diagram, but slowly inflates via a different pathway as reflected by (6) and (7). The dotted arrows denote the direction of the LU cycle. (b) Plot of $Q_{\ell}$ during LU cycle. We see significant deviation from icosahedral symmetry and the emergence of octahedral symmetry. The asymmetry of the curves about point 4 reflects the hysteretic behavior of the system. (c) Plot of $|W_{\ell}|$ during LU cycle. }
\label{fig:PV}
\end{figure}


Having considered the symmetry of the buckled shells, we now consider their mechanical response as they buckle, focusing our attention on the $(P, 0)$ icosadeltahedral shell \cite{Nelson03, Nelson07} which deforms into an icosahedral structure first. We focus on the case $(P,0) = (8,0)$, although the behavior for other values of $P$ is similar. On isotropic compression of a hollow spherical shell, it buckles and becomes faceted into an increasingly icosahedral shape, until eventually, at a scaled upper critical buckling pressure $p_b^u \approx 210$, the shell collapses abruptly into a structure with the symmetry of a squashed cube shown in Fig.~\ref{fig:PV}. A transition from an icosahedral to squashed cubic symmetry is plausible, because an icosahedron can be dissected into three orthogonal golden rectangles \cite{Coxeter}. The snap-through transition presumably selects one of five equivalent dissections and then squashes along one of the three orthogonal directions. Classical continuum elastic theory shows that the buckling pressure of an ideal sphere under hydrostatic pressure is $p_b^c = 4\sqrt{\kappa Y}/R^2 \approx 230$ \cite{Hut67}, surprisingly close to our simulation results despite differing from the continuum theory in two important ways: our shells are crystalline and have topological defects. These two features evidently partially compensate. Next, we reduce the ambient pressure and we find that the shell remains cube-like with the 6 bulges becoming less pronounced, until eventually, at a lower critical pressure $p_b^l \approx 50$, the bulges pop back out and the shell recovers its icosahedral shape (see Fig.~\ref{fig:PV}). Thus we see a strongly hysteretic transition in the morphology of these Platonic shells as a function of pressure (or volume) in our simulations. 

Quantitatively, as the shell becomes highly buckled, we see the emergence of $\ell = 4,8, \dots$ modes, typically associated with octahedral-symmetry as well as the $\ell = 2$ mode which does not belong to the icosahedral group, tetrahedral group or the octahedral group. During the snap-through transition, akin to a first order phase transition, the $\ell = 2, 4$ modes are excited as seen from Fig.~\ref{fig:PV}b and remain significant even as the pressure is reduced. The $W_{\ell}$'s highlight this effect wherein at high external pressure, we see the emergence of approximate octahedral symmetry. Indeed, at point 4, we find that $|W_{4}|  = 0.1529$, a value very close to that of the octahedron. Evidently, there is a {\it spontaneous} breaking of icosahedral symmetry during this abrupt buckling transition and an emergence of another symmetry group corresponding to a $d$-wave excitation ($\ell = 2$) mode. We find that $W_{2} \approx -0.233$ during the return portion of the hysteresis loop which also exhibits {\it return point memory} \cite{Set93}, i.e. the systems returns to the original curve at exactly the same state that it left. The negative value of $W_2$ indicates a reduced symmetry which is oblate as opposed to prolate, scaling with $Y_2^0 \propto (3 \cos^2 \theta - 1)$, averaged over all the vertices \cite{foot2}. Similar effects are seen for the large deformation behavior of other Platonic shells (see \cite{supmat}).

To understand these transitions, we note that for a featureless spherical shell with perfect symmetry, $Q_{\ell} = \delta_{\ell, 0}$. As the shell buckles, some of the $a^m_{\ell}$ for $\ell \neq 0$ will become non-zero. A shell with a broken spherical symmetry is characterized by a set of dominant $\ell$ modes, $\Lambda$, that characterizes the buckled shape 
\ba
\delta D(\theta, \phi) &= &D(\theta, \phi) - a_0^0 Y_0^0 \nonumber \\
& \approx & \sum_{\ell \in \Lambda} \sum_{m=-\ell}^{\ell} a_{\ell}^m \, Y_{\ell}^{m}(\theta, \phi) + \cdots 
\ea
This observation allows us to use a Landau-like theory of phase transitions \cite{Goshen71, Nelson81, Nelson83}, by writing down a free energy involving rotationally invariant combinations of $a_{\ell}^{m}$'s given by
\be
F = \sum_{\ell_i\in\Lambda} F_{\ell_i} + \sum_{\ell_i, \ell_j \in \Lambda, i \neq j} F_{\ell_i, \ell_j} + \cdots,
\label{eq:generalF}
\ee where 
\be
F_{\ell} = \alpha_{\ell}\sum_{m = -\ell}^{\ell} |a_{\ell}^{m}|^2+\beta_{\ell} \displaystyle{
\sum_\Omega \left( 
\begin{array}{ccc}
\ell & \ell & \ell \\ 
m_1 & m_2 & m_3 
\end{array} \right) a_{\ell}^{m_1} a_{\ell}^{m_2} a_{\ell}^{m_3} }
\label{eq:Landau1}
\ee
and 
\be
F_{\ell_i, \ell_j} =\gamma_{\ell_i, \ell_j} \displaystyle{
\sum_\Omega \left( 
\begin{array}{ccc}
\ell_i & \ell_i & \ell_j \\ 
m_1 & m_2 & m_3 
\end{array} \right) a_{\ell_i}^{m_1} a_{\ell_i}^{m_2} a_{\ell_j}^{m_3} }.
\label{eq:Fij}
\ee
Here $\alpha_{\ell}$ and $\beta_{\ell}$ are pressure-dependent parameters whose signs determine the order of the shape transition and $\gamma_{\ell_i, \ell_j}$ measures the coupling between the modes $\ell_i$ and $\ell_j$. The presence of the coupling term $F_{\ell_i, \ell_j}$ implies that nonzero $\ell_i$ spherical harmonics can generate $\ell_j$ modes if these are not already nonzero \cite{foot3}. This coupling term is unnecessary during the slow deformation phase, but is important during the abrupt collapse phase (see Fig.~\ref{fig:PV}b,c).

During the slow buckling process of the icosahedron-shell or dodecahedron-shell, a single mode free energy $F = F_6$ suffices; likewise the slow buckling of the cube-shell and octahedron-shell can be described by $F = F_4$. If the second-order coupling constant $\alpha_{\ell}(p)$ becomes nonzero with increasing pressure, then $F_{\ell}$ will be minimized by a state such that $a_{\ell}^{m} \neq 0$ where the quadratic term in Eq.~(\ref{eq:Landau1}) dominates the free energy. Furthermore, if the third-order coupling constant $\beta_{\ell} \neq 0$, Landau theory predicts that this will be a first-order transition that leads to hysteresis \cite{Nelson81, Nelson83, Goshen71}. For single mode shape transitions, if we fix the magnitude of $Q_{\ell}$ and assume the transition is weakly first order, then the preferred state can be found by minimizing the third order term in Eq.~(\ref{eq:Landau1}), with the second order term held fixed \cite{Nelson83}. In general, the form of the transition is determined by finding extrema of the symmetry invariant $W_{\ell}$ \cite{Busse75, Sattinger78,Nelson83}; more details can be found in \cite{supmat}. However, to understand the full loading cycle, the complete free energy expression as given by Eq.~(\ref{eq:generalF}) is required.  

Our analysis of the buckling process of crystalline shells with different topological defects in the frozen defect limit shows that the number, type and symmetry of the defects for shells with only one type of disclination allows us to control the resulting morphologies and generate shapes that resemble the classical Platonic solids. In general, crystalline shells with multiple types of disclinations deform into more complex structures \cite{supmat} and sites with greater topological charge have a tendency to bend/buckle more. Thus, we can generate objects with a discrete symmetry from one with a continuous symmetry using a combination of topology, geometry and mechanics. This interplay suggests a novel way to create polyhedra that differs from previous work \cite{Vernizzi11}, where the authors use two-component elastic shells, in constrast to using defects to pattern and drive the faceting transition. At a mechanical level, we have shown that a sharp buckling transition that results in spontaneous symmetry breaking leads to the appearance of pronounced spherical harmonic modes. The buckling process can be understood by studying rotational invariant quantities which quantifies the symmetry of the structure in terms of a Landau free energy model that captures the symmetry-breaking transition of the shell during the full loading and unloading cycle of the crystalline shell. Since we have focused on the zero temperature limit of the problem, our results are also applicable to macroscopic shells that are made of discrete elements and suggest a simple way to trigger shape changes between smooth and faceted structures on all scales. 

We acknowledge support from the Harvard MRSEC DMR0820484, NSF DMR1005289 (DRN), the Kavli Institute for Nano-bio Science and Technology and the MacArthur Foundation (LM).

\appendix


\section{The topology of defects}

Euler's formula is given by 
\[
\chi_E = V - E + F,
\]
where $\chi$ is the Euler characteristic and $F$, $E$, $V$ are the number of faces, edges and vertices respectively in a triangulation. In particular, $\chi_E = 2(1-g) = 2 $ for a shell (genus zero). When we triangulate the surface of a spherical manifold, since there are exactly three vertices per face and two vertices per edge, we find that $3F = 2E$. Let $N_z$ be the number of vertex with $z$-coordination number. Then $V = \sum_z N_z$ and $\sum_z z N_z = 2E$. Using these geometric relations, we find that we can rewrite Euler's formula as
\begin{equation}
\sum_z (6 - z) N_z = 6 \chi_E = 12.
\label{Euler}
\end{equation}

Thus, Euler's formula implies that it is impossible to have a crystalline shell with only 6-fold coordinated sites and if we consider a genus zero surface with only 5-, 6- and 7-fold coordinated sites, there will be exactly twelve more 5-fold than 7-fold sites due to the topology of the shell. A disclination is a lattice site with coordination other than six. 

When the core energies are large so that the creation of topological defects is heavily penalized, the spherical shell prefers to have the minimum number of disclinations that satisfy equation (\ref{Euler}). On a triangulated spherical surface, we can have 3-, 4-, 5-, 6-, 7-, 8-, 9-fold vertices. It may be useful to think of each site, with coordination number $z$, as having a topological charge of $6 - z$ (indicating the strength of the disclination), e.g. 3-fold site has charge +3,  4-fold site has charge +2 etc. Then equation (\ref{Euler}) can be understood as a conservation of topological charge statement, i.e., the sum of topological charges on a spherical crystal lattice must add up to 12. Since 3-, 4-, 5-fold defects contribute positively, while 7-, 8-, 9-fold defects contribute negatively to the summation in equation (\ref{Euler}), this means that any crystalline spherical surface must have a net excess of 3-, 4-, 5-fold defects compared to 7-, 8-, 9-fold defects in a sea of 6-fold sites that must add up to 12. Since we are looking at the minimal number of topological defects (i.e., lowest energy configurations), we can restrict our attention to combinations of 3-, 4-, 5-fold defects. Let us denote the number of 3-, 4-, 5-fold disclinations by $\{n_3, n_4, n_5\}$, e.g. a shell with 12 5-fold disclinations would be denoted by $\{ 0, 0, 12\}$. There are a total of 19 different possibilities and they fall into three distinct universality classes based on the number of distinct topological defects they have as shown in Table \ref{defect_class}.

\begin{table}[h]
\caption[Different configurations of topological defects]{\label{defect_class}Different configurations of topological defects. } 
\begin{tabular}{ll}
 \hline                       
\# distinct defects & Different cases\\
 \hline
1 & $\{ 0, 0, 12\} \Rightarrow$ icosahedron/dodecahedron, \\
&$\{ 4, 0, 0\} \Rightarrow$ tetrahedron, \\
&$\{ 0, 6, 0\} \Rightarrow$ octahedron/cube\\
\hline
2 & $\{ 0, 1, 10\}$, $\{ 0, 2, 8\}$, $\{ 0, 3, 6\}$,  \\
& $\{ 0, 4, 4\}$, $\{ 0, 5, 2\}$, $\{ 1, 0, 9\}$, \\
& $\{ 2, 0, 6\}$, $\{ 3, 0, 3\}$, $\{ 2, 3, 0\}$ \\
\hline
3 & $\{ 1, 1, 7\}$, $\{ 1, 2, 5\}$, , $\{ 1, 3, 3\}$, \\
& $\{ 1, 4, 1\}$, $\{ 2, 1, 4\}$, $\{ 2, 2, 2\}$, \\
& $\{ 3, 1, 1\}$ \\
 \hline
\end{tabular} 
\end{table}

\section{Numerical simulation algorithm}

In order to understand the morphology of a buckling thin shell, we use Surface Evolver \cite{Brakke92} to simulate the deformation of the shell under the constraint of a decreasing volume. Surface Evolver is an interactive program for the study of surfaces shaped by surface tension or other energies, and subject to various constraints. The total energy of a thin two-dimensional shell can be expressed as a sum of the stretching energy and the bending energy, i.e., $U_T = U_S + U_B$ \cite{Witten07, Landau86} and evolves toward minimal energy via a gradient or conjugate gradient descent method. For a closed surface with fixed topology, and when $\kappa_G$ is a constant, the Gaussian curvature term integrates to a constant by the Gauss-Bonnet theorem and will henceforth be dropped, as it will have no influence on the morphology of the shell. Hence the bending energy only includes the contribution from the mean curvature. Thus in our numerics, we need only consider the elastic strain from the stretching of the shell and the bending energy from mean curvature. 

The surface is implemented as a simplicial complex, that is a union of triangles, and each triangle (face) is uniquely defined by its three vertices ${\bf v}_1$, ${\bf v}_3$ and ${\bf v}_3$. Let ${\bf s}_1 = {\bf v}_2-{\bf v}_1$ and ${\bf s}_2 = {\bf v}_3-{\bf v}_1$ be the unstrained sides of the triangle, and construct column matrices $S = [{\bf s}_1, {\bf s}_2]$. When the surface is strained, the three vertices of each triangle is deformed from ${\bf v}_i$ to ${\bf v}_i'$. In a similar manner, let ${\bf r}_1 = {\bf v}_2' - {\bf v}_1'$ and ${\bf r}_2 = {\bf v}_3'-{\bf v}_1'$ be the strained sides and construct $R = [{\bf r}_1, {\bf r}_2]$. The deformation gradient matrix $D$ satisfies $DS = R$ and the Cauchy-Green strain matrix is given by $C = (D^{T}D - I)/2$, where $I$ is the $2\times2$ identity matrix. Then the strain energy density is
\be
U_S = \frac{Y}{2(1+ \nu)}\left( Tr(C^2) + \frac{\nu}{(1 - \nu)} (TrC)^2 \right), 
\ee
where $Y =  Eh$ is the two-dimensional Young's modulus and $\nu$ is the Poisson ratio. Each vertex $v$ has a star of triangles around it of total area $A_v$. The force on each vertex $v$ is 
\be
F_v = -\frac{\partial A_v}{\partial V}.
\ee
Since each triangle has 3 vertices, the area associated with $v$ is $A_v/3$. Then the mean curvature is 
\be
h_v = \frac{F_v}{2(A_v/3)}.
\ee
The bending energy density is then
\be
U_B = \kappa h_v^2 (A_v/3) = \kappa \frac{F_v^2}{4(A_v/3)},
\ee
where $\kappa= Eh^3/12(1-\nu^2)$ is the bending rigidity. The ratio of bending rigidity to 2D Young's modulus $\kappa/Y$ and the poisson ratio $\nu$ defines the thickness of the shell $h$, since for a uniform material, $h^2 = 12(1-\nu^2) \frac{\kappa}{Y}$. Our model for the undeformed spherical shell has varying thickness $h$, initial volume $V_0 = 4\pi R^3/3 = 1$ (initial radius $R_0 = 0.62$) and spontaneous mean curvature equal to $H_0 = 1/R_0 = 1.61$. The Poisson ratio is determined to be $\nu = 1/3$ \cite{Seung88, Quilliet08} and the material is assumed to be isotropic with constant elastic moduli $E$. In our simulations, we vary the 2D Young's constant $Y$ and fix $\kappa = 1$ so as to get a reasonable range of $h/R$. Starting from a initial shell, we optimize the energy to get the equilibrium shape. Then we decrease the volume at a given constant rate $\delta = \Delta V/V$ per time step until a final volume. At each volume step, Surface Evolver will converge the surface toward the minimal energy by a gradient descent method. For $\delta \ll 1$, the shell is evolving quasi-statically and is always at equilibrium. We control the volume and the pressure appears as a Lagrange multiplier. Alternatively, we may control the pressure in which case the volume would be a Lagrange multiplier.

\section{Phase space}

\begin{figure*}[htbp]
\centering
\includegraphics[width=6in]{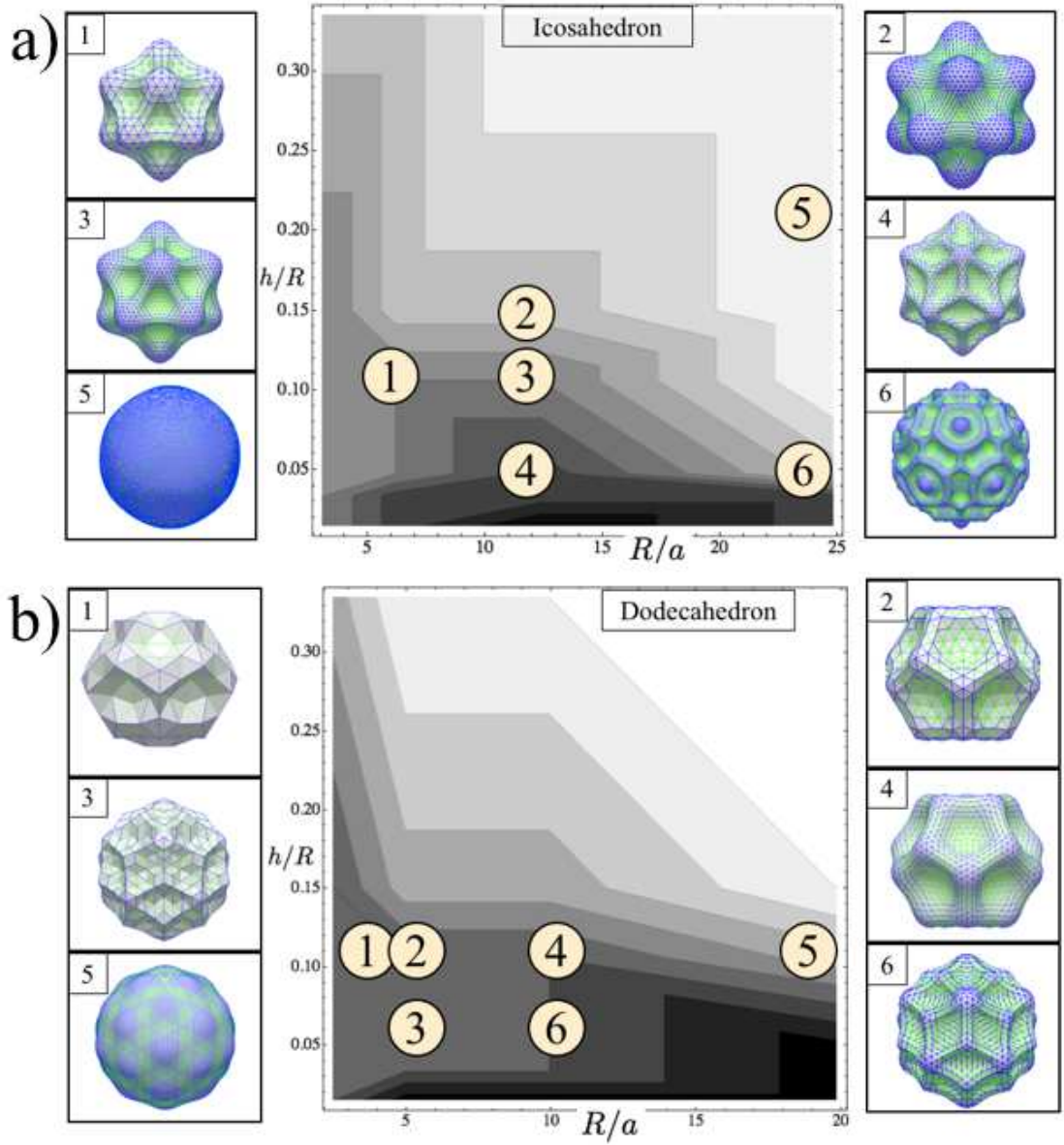}
\caption{(Color online) (a) The morphological phase diagram of shell with icosahedral defects as a function of the aspect ratio $h/R$ and lattice ratio $R/a$. (1) corresponds to a shell with $(h/R, R/a) = (0.12, 3.1)$; (2) $(0.15, 6.2)$; (3) $(0.12, 6.2)$; (4) $(0.05, 12.4)$; (5) $(0.22, 24.8)$; (6) $(0.04, 24.8)$. In all the simulations, the volume of the shell is reduced by approximatly 20$\%$. Color scheme used: White = spherical; Black = buckled; Grey = polyhedral.  (b) The morphological phase diagram of shell with dodecahedral defects. Blue (green) color indicates edges that are convex (concave).}
\label{fig:ico_dodec_space}
\end{figure*}

\begin{figure*}[htbp]
\centering
\includegraphics[width=6in]{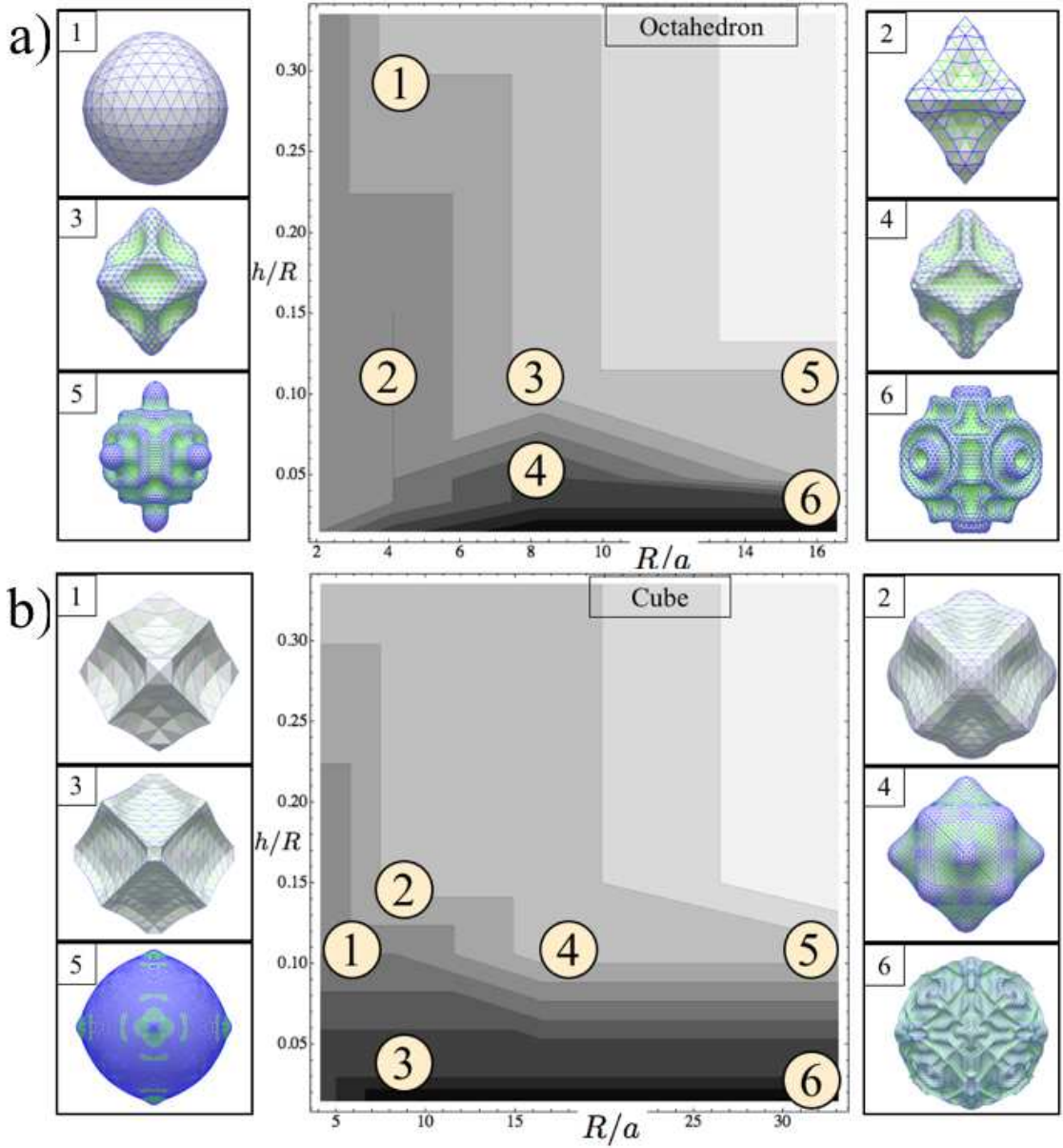}
\caption{(Color online) (a) The morphological phase diagram of shell with octahedral defects as a function of the aspect ratio $h/R$ and lattice ratio $R/a$. (b) The morphological phase diagram of shell with cubical defects as a function of the aspect ratio $h/R$ and lattice ratio $R/a$. Color scheme used: White = spherical; Black = buckled; Grey = polyhedral. Blue (green) color indicates edges that are convex (concave).}
\label{fig:octa_cube_space}
\end{figure*}

\begin{figure*}[htbp]
\centering
\includegraphics[width=6in]{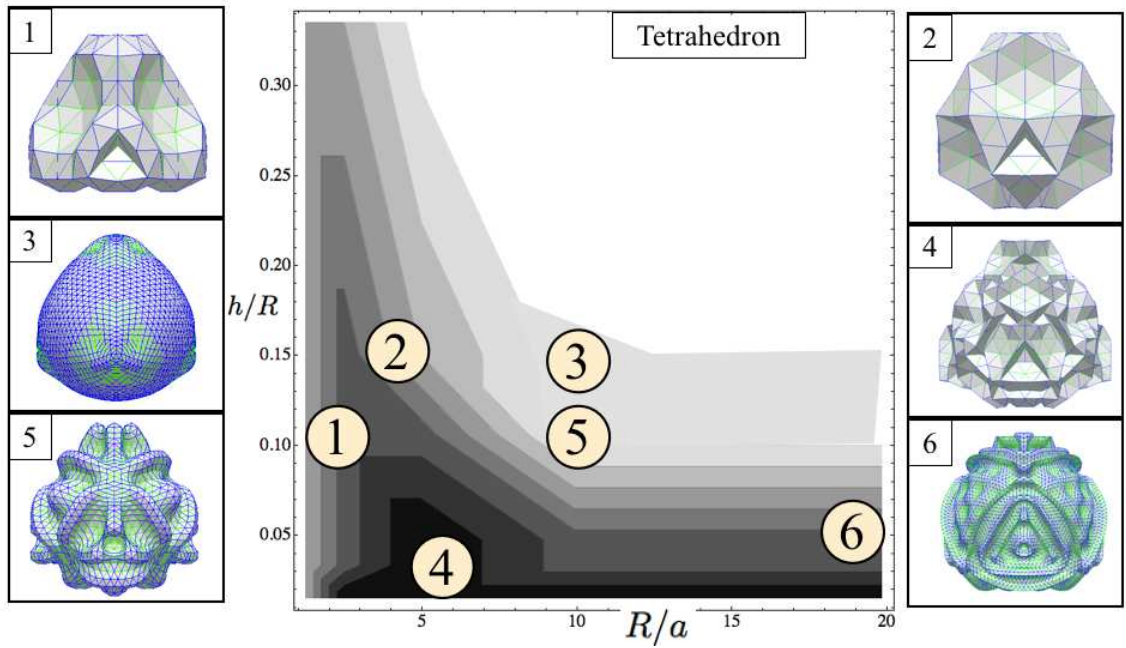}
\caption{(Color online) The morphological phase diagram of shell with tetrahedral defects as a function of the aspect ratio $h/R$ and lattice ratio $R/a$. Color scheme used: White = spherical; Black = buckled; Grey = polyhedral. Blue (green) color indicates edges that are convex (concave).}
\label{fig:tetra_space}
\end{figure*}

The shape of the deformed shell depends sensitively on its thickness $h$, its radius $R$ and its lattice spacing $a$ from which we can construct 2 dimensionless parameters, $h/R$ (aspect ratio) and $R/a$ (lattice ratio). To understand in what regions of phase space $(R/a, h/R)$ we get a buckled shell that resembles the classical Platonic shells, we change the type, number and location of the defects described earlier, and use the reduction of volume of the shells to determine the resulting shapes. The results are shown in Figs.~\ref{fig:ico_dodec_space}, \ref{fig:octa_cube_space} and \ref{fig:tetra_space}. The aspect ratio parameter $h/R$ can be varied freely since it is controlled by the two elastic modulus $Y$ and $\kappa$ that we can set arbitrarily. Due to the algorithm that we use for subdivision, in which each edge is divided in two, and each triangle is divided into four new ones in a regular manner, $R/a$ can only take on certain values at each step of the mesh refinement resulting in a somewhat coarser resolution in the $R/a$ direction in the phase space.  

\begin{figure*}[htbp]
\centering
\includegraphics[width=6in]{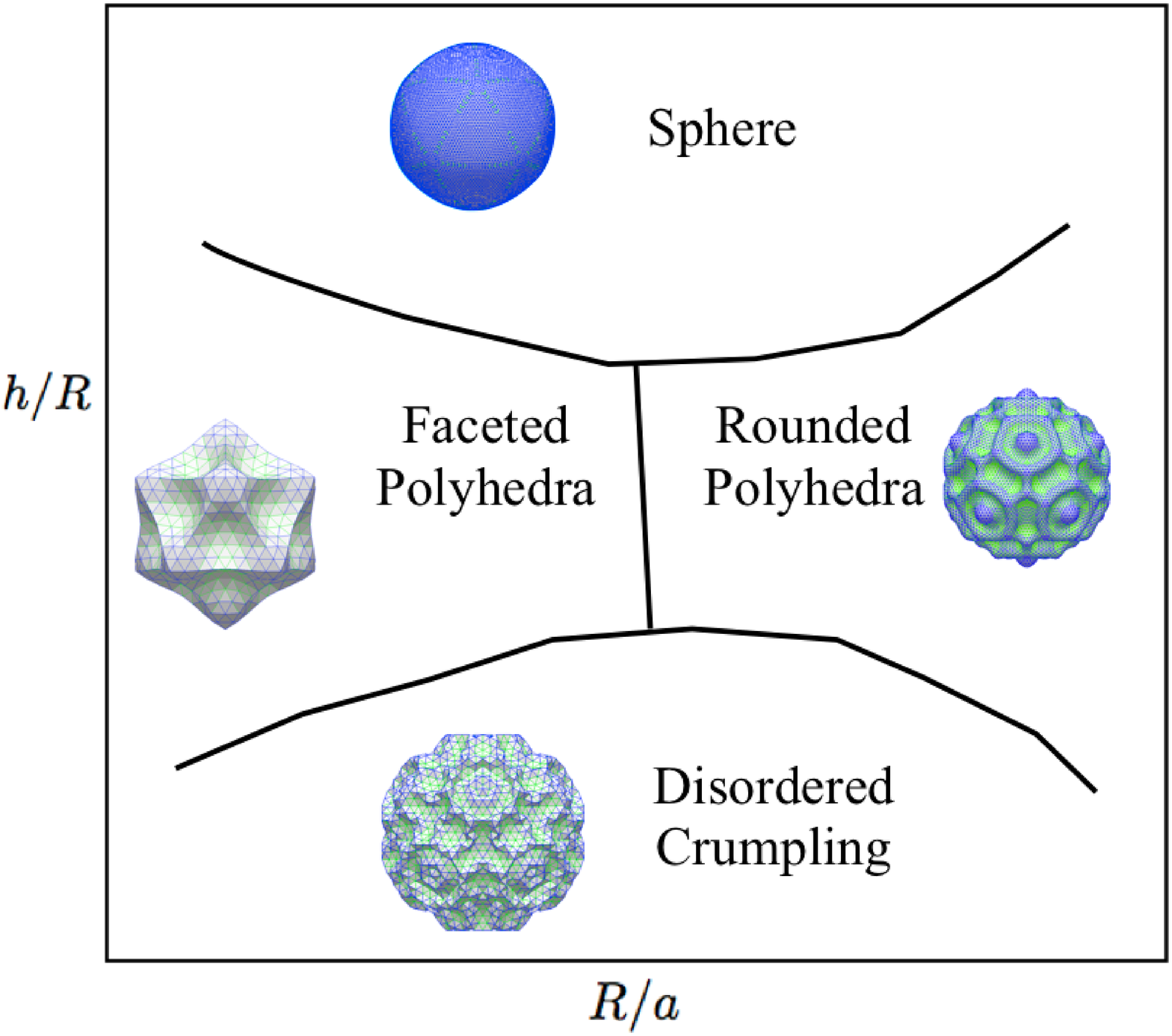}
\caption{(Color online) Phase diagram as a function of $h/R$ and $R/a$, illustrated for the icosahedral case. For small $h/R$ and $R/a$, we get simple faceted buckled final states. As $R/a$ increase, the buckled shell gets more complex. On the other hand, as $h/R$ increase, bending becomes increasingly energetically prohibitive, resulting in less buckling. }
\label{fig:psummary}
\end{figure*}

The phase diagrams for crystalline shells with different symmetries have the same qualitative features. As we vary the aspect ratio $h/R$, we find that thick shells ($h/R \gtrsim 0.1$) tend to buckle isotropically and the deformed structure is generally smooth with no noticeable faceting; thin shells tend to buckle into highly faceted structures. On the other hand, at low lattice ratio $R/a$, we get simple buckled structures since there are fewer degrees of freedom; as we increase $R/a$, we get more and more complicated deformed structures. At intermediate values of $h/R$ and $R/a$, we get structures that look like the classical Platonic polyhedra. The similarities in the five phase diagrams suggest that the buckling behavior is very generic and the phase diagram for any triangulated lattice should look qualitatively like Fig.~\ref{fig:psummary}.  

\section{Shape spectroscopy using the rotational invariant $Q_{\ell}$}

The set of spherical harmonics $Y_{\ell}^{m}(\theta, \phi)$ for a given $\ell$ forms a $(2\ell + 1)$-dimensional representation of the rotational group SO(3). This means that the $a_{\ell}^{m}$'s for a given $\ell$ can be scrambled by a simple rotation of coordinates via \cite{MW1970}
\be
(a_{\ell}^{m})_{\textrm{new}}= \sum_{m_1} D_{m, m_1}^{(\ell)}(\alpha, \beta, \gamma) \,a_{\ell}^{m_1}.
\ee
Therefore, it is better to look at rotationally invariant quantities formed from the various $a_{\ell}^{m}$ that measure the angular projection onto the different $\ell$, such as the invariant quantity $q_{\ell}$ \cite{Nelson83}, defined as
\be
q_{\ell} = \left( \frac{4 \pi}{2\ell +1} \sum_{m = -\ell}^{\ell} |a_{\ell}^m|^2\right)^{1/2}.
\ee
$q_{\ell}$ measures the ``magnitude" of the $(2\ell + 1)$ order parameter $a_{\ell}^{m}$. The quantity $a_0^0$, which corresponds to a constant spherical harmonics $Y^0_0 = 1/\sqrt{4\pi}$, is always nonzero, and scales with the size of the initial shell $R$. Therefore, it is useful to normalized the various $q_{\ell}$ by $q_0$, so that it is independent of the overall magnitude of the $\{a_{\ell}^m\}$ for any given $\ell$, namely,
\be
Q_{\ell} = q_{\ell}/q_0.
\label{eq:Qell}
\ee
By definition $Q_{0}$ = 1. A large value of $Q_{\ell}$ for $\ell \neq 0$ implies a great degree of aspherity.

The initial shell has perfect spherical symmetry and the only nonzero mode is the $\ell = 0$ mode, i.e., only $a_0^0$ is non-vanishing. However, as the shell buckles, the deformed shell starts to take on interesting shapes, with nonvanishing $Q_{\ell}$'s, where $\ell \neq 0$. For example, $Q_{3}$ measures tetrahedral symmetry, $Q_{4}$ measures tetrahedral and cubic symmetry and $Q_{6}$ measures icosahedral symmetry and so on. We calculate the $Q_{\ell}$'s for the different Platonic-shells at volume reduction of around $20\%$. We find that the buckled icosahedral-shell has nonzero spherical harmonics only for $\ell = 0, 6, 10, 12, ...$, which strongly suggests that the set of topological defects is able to enforce the icosahedral symmetry as it buckles. The dodecahedron-shell has the same angular momentum $\ell$ modes as the icosahedron-shell. The cube-shell and octahedron-shell have nonvanishing spherical harmonics for $\ell = 0, 4, 6, 8, 10, ...$ and the tetrahedron-shell has $\ell =  0, 3, 4, 6, 7, 8, 9, 10, ...$. Some typical values of $Q_{\ell}$ for the Platonic-shells with $h/R = 0.118$ are shown in table~\ref{tab_Q}.
\begin{table*}[htbp]
\caption{\label{tab_Q} Normalized invariant $Q_{\ell}$ for the different Platonic-shells with $h/R = 0.118$. $Q_2 = Q_5 = 0$ for all Platonic shells. }
\begin{tabular}{l|l|lllllll}
\hline                        
Type of shell & $R/a$ & $Q_{3}$ & $Q_{4}$ & $Q_{6}$ & $Q_{7}$ & $Q_{8}$ & $Q_{9}$ & $Q_{10}$ \\ 
\hline
Icosahedron & 10.9 &  0 & 0 &  0.0351 & 0 & 0 & 0 & 0.0093 \\
Dodecahedron & 9.5 &  0 & 0 & 0.0142 & 0 & 0 & 0 & 0.0170 \\
Cube & 12.3 & 0 &  0.0235 & 0.0126 & 0 & 0.00629 & 0 &  0.0099 \\
Octahedron & 7.3 &  0 & 0.0412 &  0.0492 & 0 &  0.00030 & 0 & 0.0098 \\
Tetrahedron &  9.7 &  0.0287 & 0.0251 & 0.0030 & 0.0047 & 0.00034 & 0.00234 & 0.0009 \\
\hline
\end{tabular}
\end{table*}

The normalized invariant $Q_{\ell}$ varies as we change the two non-dimensional parameters $h/R$ as well as $R/a$. The values of the normalized $Q_{\ell}$ are illustrated in Table~\ref{QL}, which shows $Q_{6}$ and $Q_{10}$ for different values of $h/R$ and $R/a$ for the icosahedral-shell. As $R/a$ increases, we find that $Q_{6}$ decreases while $Q_{10}$ increases, which implies that higher order $\ell$ modes becomes more important as the mesh spacing becomes finer, which is very reasonable. Similarly, we find that as the shell becomes thinner (smaller $h/R$), $Q_{6}$ decreases while $Q_{10}$ increases since more excitations are transferred from the lower order $\ell$ modes to the higher order $\ell$ modes.   

\begin{table}[htbp]
\caption{\label{QL} Normalized invariant $Q_{6}$ and $Q_{10}$ for different values of $h/R$ and $R/a$ of the icosahedron-shell.}
\begin{tabular}{l|l|ll||l|l|ll}
\hline                            
$h/R$ & $R/a$ & $Q_{6}$ & $Q_{10}$ & $h/R$ & $R/a$ & $Q_{6}$ & $Q_{10}$ \\
\hline
0.235 & $5.55$ &$0.0363$ & $0.0095$ &0.0745 & $5.55$ &$0.0177$ & $0.0150$\\
0.235 & $11.55$ &$0.0080$ & $0.0003$ &0.0745 & $11.55 $ &$0.0200$ & $0.0155$\\
0.235 & $22.87$ &$0.0039$ & $0.0012$ &0.0745 & $22.87 $ &$0.00580$ & $0.00001$\\
0.118 & $5.55$ &$0.0363$ & $0.0095$ & 0.0372 & $5.55$ &$0.00583$ & $0.00004$\\
0.118 & $11.55$ &$0.0357$ & $0.0097$ & 0.0372 & $11.55 $ &$0.00819$ & $0.0199$\\
0.118 & $22.87$ &$0.0038$ & $0.0170$ & 0.0372 & $22.87 $ &$0.00517$ & $0.0014$\\
\hline
\end{tabular}
\end{table}

\section{Mechanical response of crystalline shells with different symmetries}

The case of the icosahedron-shell is described in the main text. Here, we summarize results for the remaining 4 Platonic-shells: dodecahedron-shell, cube-shell, octahedron-shell and the tetrahedron-shell. 

\subsection{Dodecahedron-shell}

\begin{figure*}[htbp]
\centering
\includegraphics[width=6in]{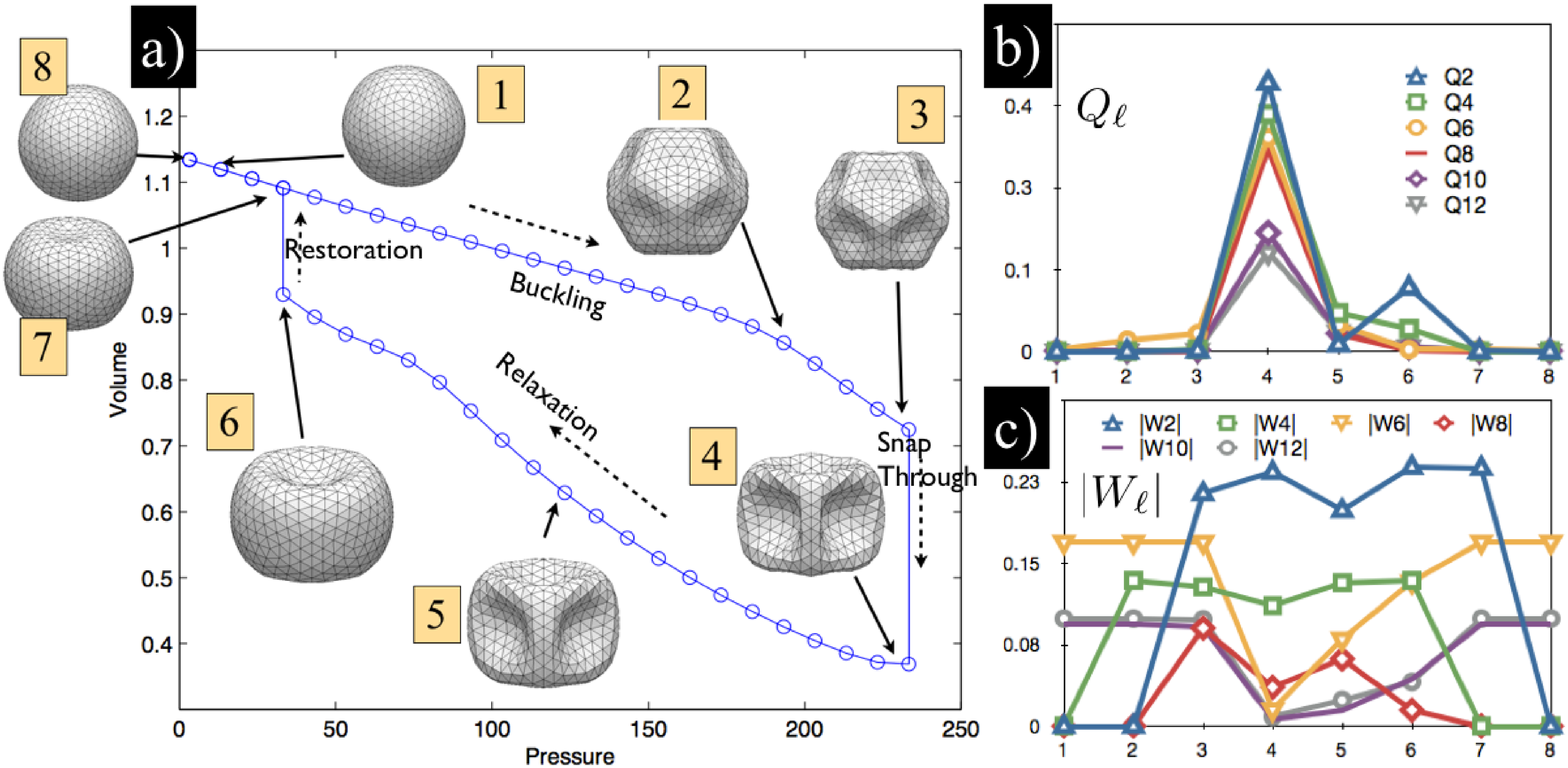}
\caption{(Color online) (a) Mechanical response of a crushed dodecahedron-shell. The labels $(1)-(8)$ show the shape of the shell at different points of the loading-unloading cycle. (b) Plot of $Q_{\ell}$ of the dodecahedral-shell during LU cycle. (c) Plot of $|W_{\ell}|$ during LU cycle. }
\label{fig:dodec_LU}
\end{figure*}

The buckling behavior of the dodecahedron-shell (also known as the $(P, P)$ icosadeltahedral shell) is similar to that of the icosahedron-shell. In our case, we consider $P = 4$, which has 482 vertices, 1440 edges and 960 facets, with twelve 5-fold disclinations. The initial spherical shell quickly becomes faceted at the locations of the twelve symmetrically positioned five-fold coordinated points of the triangular mesh resulting in a structure that resembles a dodecahedron. As we increase the pressure, the facets of the shell become concave inward, and we see the emergence of twelve inward bulges as shown in Fig.~\ref{fig:dodec_LU}. These dimples become more pronounced as we increase the pressure until at a critical buckling pressure $p_b \sim 230$, the shell loses stability and collapses, forming a structure that looks like a slightly squashed cube (label (4)). Thus, the number of inward bulges reduces from twelve to six and they become deeper as the pressure is increased even further. As the pressure is decreased, the bulges become shallower until eventually four of the bulges disappear and the shell forms a discocyte-like structure. It is axisymmetrical with two inward dimples at the north and south poles (label (6)). As the pressure is decreased still further to the original pressure, the shell resumes its original form as shown as shown in label (7) of Fig.~\ref{fig:dodec_LU}. Consistent with the higher buckling pressure $p_b$, the hysteresis is larger for the dodecahedral-shell compared to the icosahedral-shell. 

Our shape analysis of the dodecahedron-shell under hydrostatic pressure reveals a similar picture as compared to the icosahedral-shell as illustrated in Fig.~\ref{fig:dodec_LU}. During the buckling phase, there is a spontaneous breaking of the icosahedral symmetry and the emergence of additional $\ell$ modes. As evident from Fig.~\ref{fig:dodec_LU}, the $Q_\ell$'s for $\ell = 2,4,6,8$ modes become comparable ($\sim$30\%) to $\ell = 0$ mode. The slight asymmetry of the curves about shape (4) reflects the hysteretic behavior of the system. For the shape indicated by the label (6), we see that there is a spike in $\ell = 2$ mode. Initially, the nonvanishing $W_\ell$'s are $\ell = 6, 10,12$. As the shell buckles, $|W_{4}|$ increases from 0 to $0.13$, suggesting the emergence of octahedral/tetrahedral-symmetry. The value of $|W_{4}|$ stays pretty constant at 0.13 along the hysteresis loop except at the start and end points where it almost vanishes. Also, we see that during the buckling transition, $|W_{6}|$ decreases from $0.17$ to $0.015$ indicating a suppression of icosahedral-symmetry ($|W_{6}^{ico}| = 0.17$) and the emergence of the octahedral-symmetry ($|W_{6}^{cube}| = 0.013$). $|W_{8}|$ stays around 0.05 during buckling, which is fairly close to that of the octahedral/tetrahedral symmetry. Additionally, $W_{2} \approx +0.23$ after buckling, indicating the presence of $d$-wave excitations. Unlike the icosahedron-shell, in this case, the positive value of $W_2$ indicates a reduced symmetry which is prolate in nature. Overall, we find that the dodecahedron-shell displays similar hysteretic behavior to the icosahedral-shell.  

\subsection{Cube-shell and octahedron-shell}

\begin{figure*}[htbp]
\centering
\includegraphics[width=6in]{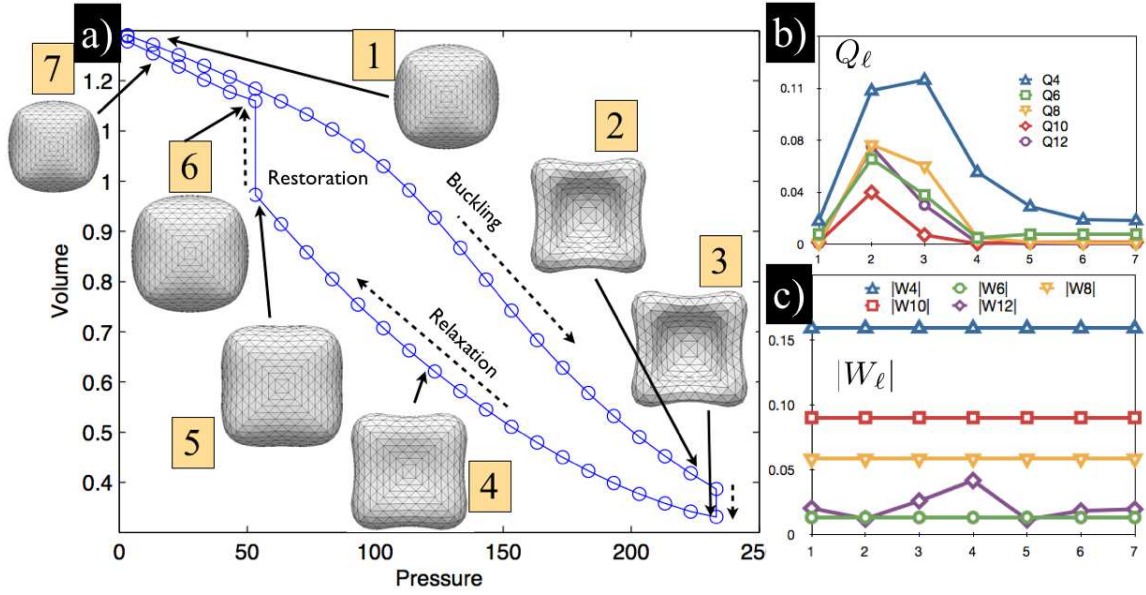}
\caption{(Color online) (a) Mechanical response of a crushed cube-shell during loading-unloading (LU) cycle. (b) Plot of $Q_{\ell}$ during LU cycle. (c) Plot of $|W_{\ell}|$ during LU cycle.}
\label{fig:cube_LU}
\end{figure*}

For the cube-shell, denoted $(P, Q) = (8,8)$ in the Caspar-Klug notation, we have 770 vertices, 2304 edges and 1536 facets, with six 4-fold disclinations. Our simulations show that the initial spherical surface quickly becomes cubical. As we increase the pressure, the shell becomes more and more faceted, with well-defined ridges, and the buckled shape becomes more and more cubical. The six evenly distributed inward bulges become deeper as we increase the pressure resulting in a more shriveled structure. As we decrease the pressure, the shell starts to swell up again, along a different pathway. At a certain pressure, $p \sim 50$,  the bulges disappear and the shell becomes globally convex and there is a noticeably jump in the volume. We find that $Q_2 \approx 0$ throughout the LU cycle. This process is summarized in Fig.~\ref{fig:cube_LU}. The hysteresis of the cycle, defined as the difference between the maximum and minimum pressures that define the loop, is larger than that for the icosahedral and dodecahedral shells. There are three distinctive parts to the hysteresis loop: 1) the buckling phase from label (1) to (3); 2) the relaxation phase from shape (3) to (5) and the restoration (or pop-out) phase from label (5) to (6). There is no noticeable ``snap-through" phase as structurally, the shell appears to retain cubical symmetry throughout the cycle. 

The plots of $Q_{\ell}$ and $|W_{\ell}|$ reveal a similar picture as shown in Fig.~\ref{fig:cube_LU}a and b. During the buckling phase, the $Q_{4}$, $Q_{6}$, $Q_{8}$, $Q_{10}$ and $Q_{12}$ all become more prominent, while modes with other $\ell$'s are suppressed. The most important mode corresponds to $\ell = 4$, with an amplitude that is almost twice as large as the second highest mode $\ell = 8$. However, as we decrease the pressure, all the excited modes amplitudes decrease in magnitude, as the shell resumes a more spherical shape. Throughout the hysteresis cycle $|W_{\ell}|$ for $\ell = 4, 6, 8, 10$ remains constant and only $|W_{12}|$ shows a slight departure from the initial value, while the other $|W_{\ell}$'s are vanishingly small. Thus, the shrinking shell retains its cubic symmetry to a high extent through the buckling process. Hence, there is no symmetry group breaking wherein the $\ell = 2$ terms appearing during the LU cycle, e.g. $Q_2 = W_2 = 0$.
\begin{figure*}[htbp]
\centering
\includegraphics[width=6in]{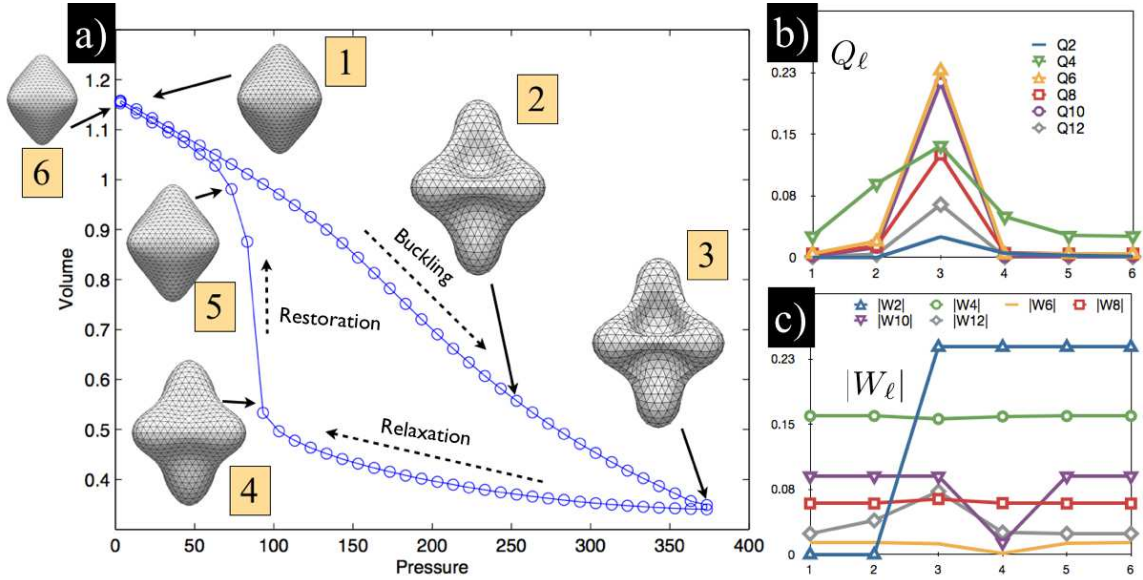}
\caption{(Color online) (a) Mechanical response of a crushed octahedron-shell during loading-unloading (LU) cycle. (b) Plot of $Q_{\ell}$ during LU cycle. (c) Plot of $|W_{\ell}|$ during LU cycle.}
\label{fig:octa_LU}
\end{figure*}

The case for the octahedron-shell is very similar to the cube-shell, which is unsurprising since they are dual to one another. We study a $(P,Q) = (16,0)$ in the Caspar-Klug notation, a triangulation that has 1026 vertices, 3072 edges and 2048 facets, with six 4-fold disclinations. The spherical shell quickly becomes octahedral-like in shape at the start of the simulation, consistent with what we observed in the previous section. As we increase the pressure, the shell retains the octahedral symmetry, and we see eight inward triangular bulges appear symmetrically on the surface. As the pressure is further increased, the bulges become deeper, facilitating a smaller volume. There is no noticeable buckling behavior as the shell remains octahedral-like throughout. As we decrease pressure to go the opposite sense in the hysteresis loop, the shell starts to swell up and the depressions become shallower and at $p \sim 80$, the depressions disappear as the volume jump significantly and all the concave regions disappear from the shell. The whole hysteresis loop is summarized in Fig.~\ref{fig:octa_LU}. In terms of the $Q_{\ell}$, we find that the most important modes are $\ell = 6$ and $10$ followed by $\ell = 4$ and $8$. We do see $\ell = 2$ mode appear although its effect appear to be relatively small compared to the other modes as seen in Fig.~\ref{fig:octa_LU}b. For the case of $|W_{\ell}|$, we find that $|W_{4}|$, $|W_{6}|$ and $|W_{8}|$ remain fairly constant during the whole hysteresis cycle. As the shell becomes buckled, we see the simultaneous emergence of $W_{2} \approx +0.225$, signaling a reduced symmetry which is prolate in nature and the reduction in $|W_{10}|$ and $|W_{12}|$. Overall, we see small deviation from octahedral-symmetry and the emergence of $d$-wave modes ($\ell$ = 2). 

Upon comparing the mechanical response of the octahedron-shell and cube-shell, we see that qualitatively they are very similar as both types of shells maintain high degree of octahedral-symmetry throughout the hysteresis cycle as seen from the constant values in $|W_{\ell}|$ (See Fig.~\ref{fig:cube_LU} and \ref{fig:octa_LU}). The only significant deviation is that for the octahedron-shell, we see the emergence of $\ell = 2$ mode, which is absent in the cube-shell. The hysteresis effect in both octahedron-shell and cube-shell is more pronounced as compared to the icosahedron-shell and dodecahedron-shell.

\subsection{Tetrahedron-shell}

\begin{figure*}[htbp]
\centering
\includegraphics[width=6in]{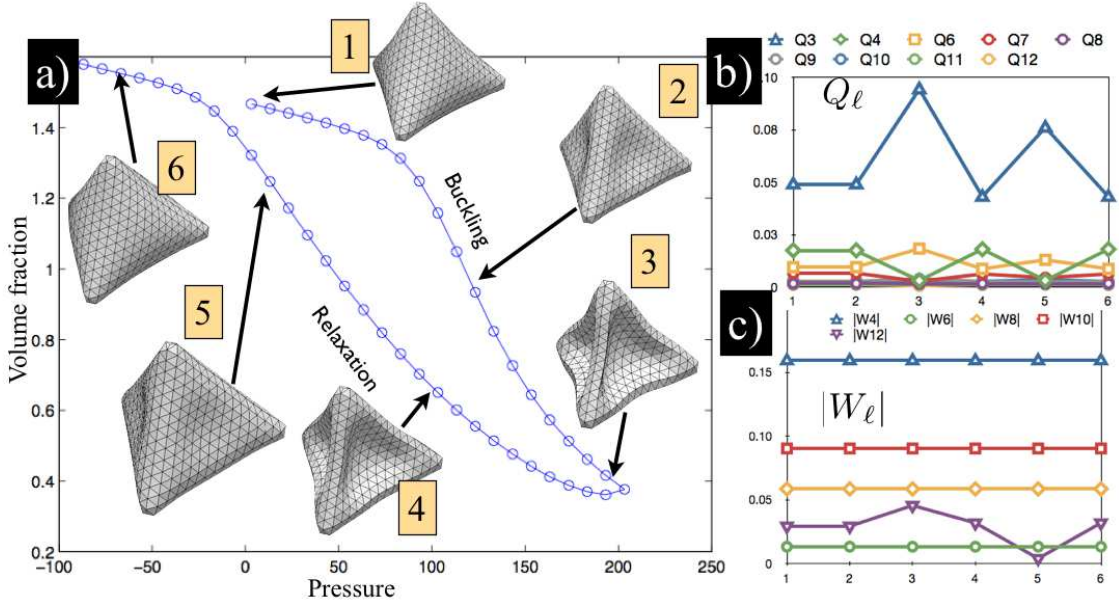}
\caption{(Color online) (a) Mechanical response of a crushed tetrahedron-shell during loading-unloading (LU) cycle. (b) Plot of $Q_{\ell}$ during LU cycle. (c) Plot of $|W_{\ell}|$ during LU cycle.}
\label{fig:tetra_LU}
\end{figure*}

Finally, we consider the $(P, Q) = (16,0)$ tetrahedron-shell. Our triangulation has 514 vertices, 1536 edges and 1024 facets, with four 3-fold disclinations. On increasing the ambient pressure, the spherical shell quickly becomes tetrahedral-like in shape at the start of the simulation due to the presence of the disclinations. As we increase the pressure, the shell retains the tetrahedral symmetry, and we see four inward triangular bulges appear symmetrically on the surface. As the pressure is further increased, the bulges become deeper, facilitating a smaller volume. There is no noticeable buckling behavior as the shell remains tetrahedral-like throughout. As the pressure is decreased, the shell starts to reswell and the dimples become shallower; eventually for $p \sim 0$, they disappear. The whole hysteresis loop is summarized in Fig.~\ref{fig:tetra_LU}. In terms of the $Q_{\ell}$, we find that the most important modes correspond to $\ell = 3$ followed by $\ell = 4$ and $6$. The other non-zero modes correspond to $\ell = 7, 8, 9, 10, 11, 12$ which are all very small in magnitude. During the loading and unloading cycle, we find that the magnitude of $Q_3$ fluctuates while the rest of the $Q$'s remain fairly constant in magnitude. In terms of the $|W_\ell|$, we find that the non-zero modes are $\ell = 4, 6, 8, 10, 12$ with $\ell = 4$ being the most dominant. The shape parameter $W_3 = 0$ throughout, even though $Q_3 \neq 0$. Presumably, this is due to the fact that the tetrahedron is self-dual. All the $|W_{\ell}|$ except for $|W_{12}|$ remain constant during the whole loading-unloading cycle implying that the symmetry is conserved during the whole loading-unloading cycle. In this case we do not find symmetry breaking although we still see the presence of hysteresis. 


\begin{thebibliography}{}

\bibitem{Taniguchi08}
M. Yanagisawa, M. Imai, and T. Taniguchi, Phys. Rev. Lett. {\bf 100}, 148102 (2008)

\bibitem{Pleiner10}
H. Pleiner, Phys. Rev. A {\bf 42}, 6060 (1990).

\bibitem{Eleni10}
E. Katifori, S. Alben, E. Cerda, D. R. Nelson, and J. Dumials, Proc. Natl. Acad. Sci. USA {\bf 107}, 7635 (2010).

\bibitem{Stone07}
M. Abkarian, A. B. Subramaniam, S.-H. Kim, R. J. Larsen, S.-M. Yang, and H. A. Stone, Phys. Rev. Lett. {\bf 99}, 188301 (2007).

\bibitem{Maha2010}
L. Mahadevan, R. Bendick, and H. Liang, Tectonics {\bf 29}, TC6002 (2010).

\bibitem{Seung88}
H. S. Seung and D. R. Nelson, Phys. Rev. A {\bf 38}, 1005 (1988).

\bibitem{Caspar62}
D. L. Caspar and A. Klug, Cold Spring Harb Symp Quant. Bio. {\bf 27}, 1 (1962).

\bibitem{Nelson03}
J. Lidmar, L. Mirny, and D. R. Nelson, Phys. Rev. E {\bf 68}, 051910 (2003).

\bibitem{Nelson07}
M. Widom, J. Lidmar, and D. R. Nelson, Phys. Rev. E {\bf 76}, 031911 (2007).

\bibitem{Datta2010}
S. S. Datta, H. C. Shum, and D. A. Weitz, Langmuir {\bf 26}, 18612 (2010).

\bibitem{Bowick00}
M. J. Bowick, D. R. Nelson, and A. Travesset, Phys. Rev. B {\bf 62}, 8738 (2000).

\bibitem{Bowick07}
M. J. Bowick, D. R. Nelson, and H. Shin, Phys Chem Chem Phys. {\bf 9}, 6304 (2007).

\bibitem{Luca07}
L. Giomi and M. Bowick, Phys. Rev. B {\bf 76}, 054106 (2007).

\bibitem{supmat}
See Appendix for more information.

\bibitem{Siber06}
A. Siber, Phys. Rev. E {\bf 73}, 061915 (2006).

\bibitem{Landau86}
L. D. Landau, L. P. Pitaevskii, E. M. Lifshitz, and A. M. Kosevich, {\it Theory of Elasticity}, 3rd ed. (Pergamon, NY, 1986).

\bibitem{Witten07}
T. A. Witten, Rev. Mod. Phys. {\bf 79}, 643 (2007).

\bibitem{Brakke92}
K. Brakke, Exp. Math. {\bf 1}, 141 (1992).

\bibitem{Nelson81}
D. R. Nelson and J. Toner, Phys. Rev. B {\bf 24}, 363 (1981). 

\bibitem{Nelson83}
P. J. Steinhardt, D. R. Nelson, and M. Ronchetti, Phys. Rev. B {\bf 28}, 784 (1983).

\bibitem{Busse75}
F. H. Busse, J. Fluid Mech. {\bf 72}, 67 (1975).

\bibitem{Sattinger78}
D. H. Sattinger, J. Math. Phys. {\bf 19}, 1720 (1978).

\bibitem{Onaka06}
S. Onaka, Philos. Mag. Lett. {\bf 86}, 175 (2006).

\bibitem{foot1}
Note from Table I that the sign of $W_\ell$ for the smallest $\ell$ for which $W_\ell$ is nonzero distinguishes the icosahedron from its dual, the dodecahedron and the octohedron from its dual the cube. For tetrahedron $W_3 = 0$, presumably because this object is self-dual.

\bibitem{Coxeter}
H. S. M. Coxeter, {\it Introduction to Geometry}, 2nd ed. (Wiley Classics Library, New York, 1999).

\bibitem{Hut67}
J. W. Hutchinson, J. Appl. Mech. , {\bf 49} (1967).

\bibitem{Set93}
J. P. Sethna, K. Dahmen, S. Kartha, J. A. Krumhansl, B. W. Roberts, and J. D. Shore, Phys. Rev. Lett. {\bf 70}, 3347 (1993).

\bibitem{foot2}
See, e.g., P. G. de Gennes and J. Prost, {\it The Physics of Liquid Crystals} (Oxford, Clarendon Press, 1993)

\bibitem{Goshen71}
S. Goshen, D. Mukamel, and S. Shtrikman, Solid State Commun. {\bf 9}, 649 (1971).

\bibitem{foot3}
Note that any discrete triangulation punctuated by a regular array of defects will inevitably lead to small nonzero values of $\{a_{\ell}^m\}$ with the appropriate symmetry, even though the overall shape is nearly spherical. However, the effect of these small background ordering fields is negligible, in light of the large first-order-like transition displayed in Fig.~3.

\bibitem{Vernizzi11}
G. Vernizzi, R. Sknepnek, and M. O. de la Cruz, Proc. Natl. Acad. Sci. USA {\bf 108}, 4292 (2011).


\bibitem{Quilliet08}
C. Quilliet, C. Zoldesi, C. Riera, A. van Blaaderen, and A. Imhof, Eur. Phys. J E {\bf 27}, 13 (2008).

\bibitem{MW1970}
J. Mathews and R. L. Walker, {\it Mathematical Methods of Physics}, 2nd ed. (W. A. Benjamin, Menlo Park, CA, 1970).


\end{thebibliography}

\end{document}